# Increasing the Efficiency of Photovoltaic Systems by Using Maximum Power Point Tracking (MPPT)

Alireza Tofigh Rihani[1], Majid Ghandchi[2]

[1]Department of Electrical Engineering, Ahar Branch, Islamic Azad University, Ahar, Iran
alireza.reahani@gmail.com (Corresponding author)
[2]Department of Engineering, Ahar Branch, Islamic Azad University, Ahar, Iran
majid.ghandchi@gmail.com

## Abstract

*Using Photovoltaic systems is gradually expanded by increasing energy demand. Abundance and availability of this energy, has turned to one of the most important sources of renewable energy. Unfortunately, photovoltaic systems have two big problems: first, those have very low energy conversion efficiency (in act between 12 and 42 percent under certain circumstances). Second, the power produced by the solar cell depends on nonlinear conditions such as solar radiation, temperature and charge feature. According to this, received power maximum of photovoltaic cells depends on different non-linear variables, it is necessary to be continuously traced, as maximum received power of the cell (by controller). In this research, the increasing efficiency of photovoltaic systems has been investigated by using Maximum Power Point Tracking (MPPT) in two different modes contained connected to the Grid and disconnected from the grid with simulation by MATLAB software. The obtained results showed that the proposed technique is able to improve the current, voltage and power output of photovoltaic cells.*

## 1- Introduction

Different MPPT algorithms can be categorized into two groups depending on offline and online models implemented. The different methods of offline and online have been investigated in [1]. Offline methods have been estimated to control signals based on primary datum and input variables of solar panel such as light intensity, temperature, open circuit voltage, short-circuit, and during operation of the system, the control signal used to control the output power panel. Open circuit voltage method [2] and [3] is one of the simplest methods of offline presented, based on approximately linear relationship between open circuit voltage maximum power voltage in different weather conditions.

$$V_{mpp} \approx kV_{oc} \qquad (1)$$

Where K is constant value, smaller than one (between 0.7 and 0.8), and depends on solar panel empirically determined by measuring the quantity and different weather conditions. In this method, the Approximate amount of $V_{mpp}$ can be measured by using equation 1 and knowing $V_{oc}$ and by setting the output voltage of the solar panel in $V_{mpp}$, the maximum can be received. In this method, we measure $V_{oc}$ value by separating the load from solar panel. While the implementation of this method is very simple and cheap but not optimized, and solar panels intermittent cut of load for calculating $V_{oc}$ is the disadvantages of this method. Short-circuit current method [4] uses approximate linear relationship between the short-circuit current $I_{sc}$ and the maximum power $I_{MPP}$ (K is between 0.8 and 0.9).

$$I_{MPP} \approx K I_{sc} \qquad (2)$$

In this method, $I_{sc}$ has been determined and then $I_{MPP}$ value has been calculated by using (2) and totally by adjusting the output current panel of this amount, through which the maximum power has been traced. By determining $I_{sc}$, the solar panels, separated from load and measure the amount of short circuit current by shortening circuit output. This method is more accurate and efficient method of open circuit voltage [4], but short-circuits current calculation problems and the cost of implementing are the disadvantages of this method. Using neural network is the other method of offline for tracking the maximum power, primarily investigated by Hiyama and et.al [5]. The open circuit voltage of the solar panel is the only neural network input data and by assistance of PI contro,ller obtained control signal is required for MPPT. In online method, usually instantaneous values of variables of solar panel (voltage and current), are used to generate control signals. As a result, control signals of these methods, unlike the offline method, are not constant and steady state oscillated around its optimal value. Perturbation and Observation method (P&O) [7] and [8], ripples convergent control method (RCC) [11] and increasing impedance method (Inc Cond) [9] including online methods are the maximum power point tracking and using output voltage and current of solar panels produce appropriate control signals.

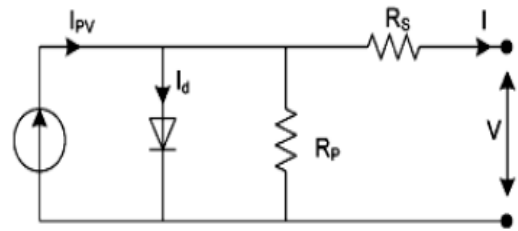

**Fig.1.** The equivalent circuit of a solar panel [15]

## 2- Photovoltaic System

Photovoltaic system consists of solar panel, battery, DC-DC voltage convertor and controller. Photovoltaic systems act like a real power supply and have a fixed voltage level of the battery used in various loads. In figure 4, battery of voltage, converter output voltage DC-DC $(V_L)$ holds constant value. Battery power is also needed for the storage and recovery point. DC-DC voltage convertors are used to match the load characteristics with characteristic of solar panels [17]. DC-DC voltage convertors are classified into three groups of: Boost, Buck and Buck-Boost convertors [18]. Selecting the type of DC-DC voltage converter depends on the level of voltage changes. In this research, DC-DC increased voltage converter using accommodating load with solar panels.

$$\frac{V_L}{V_{PV}} = \frac{1}{1-D} \qquad (3)$$

$$\rightarrow V_{PV} = (1-D)V_L \qquad (4)$$

$$if : V_{PV} = V_{MPP} \rightarrow D = D_{MPP} \qquad (5)$$

The above equations show that by changing, the operating point of the solar panel D value could be varied, and due to the uniqueness of $V_{MPP}$, there is a unique value for that point in MPP puts solar panel work. Photovoltaic system will be optimized when it has the most efficiency that solar panels can produce at any time, which is the maximum amount of the work done in MPP. By changing the switching period (D), the characteristic time for changes to D will improve the visibility of solar panels. The simulation model of Photovoltaic system is shown in figure 4.

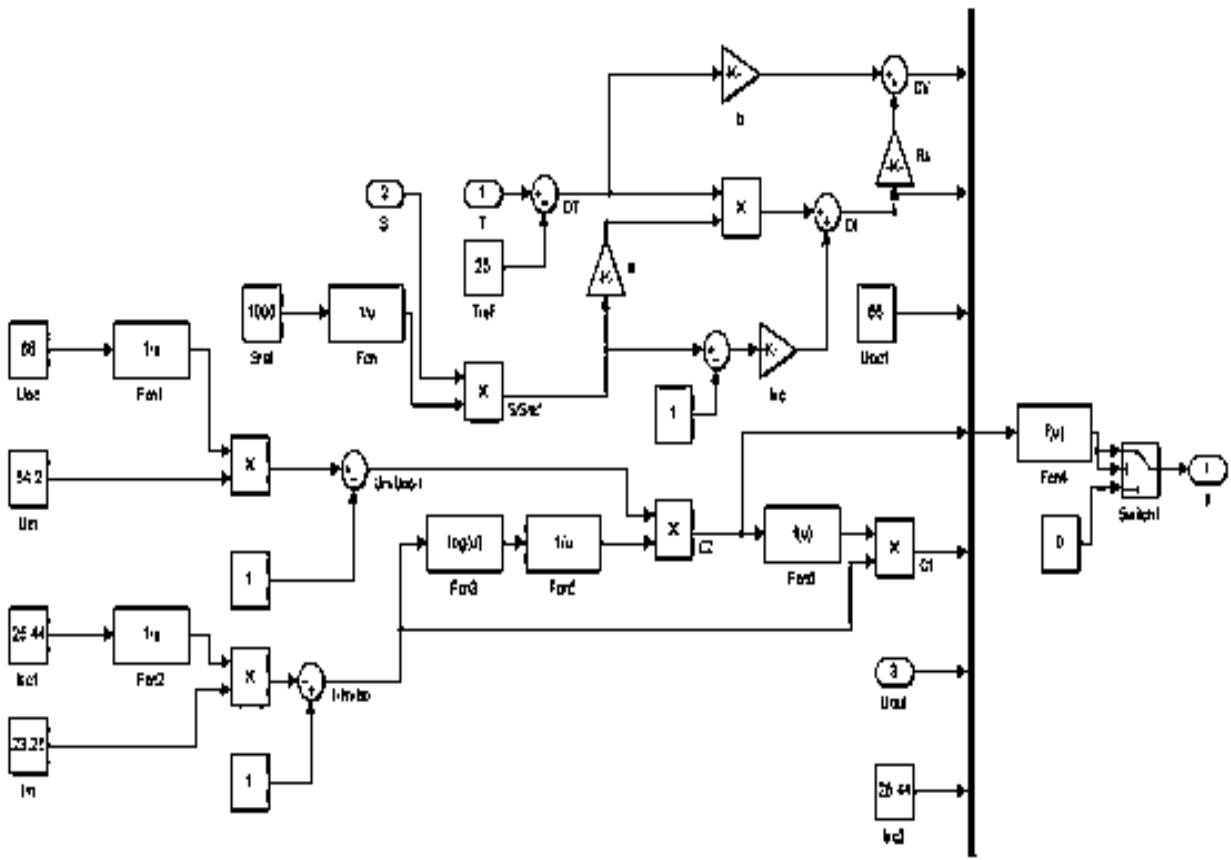

**Fig.2.** Photovoltaic system

## 3- The Proposed Method

General algorithm of proposed method is made up of two loops point calculation and precise adjustment, shown in figure 5. In the first ring, approximate amount of the maximum power point estimated and the second ring the exact amount is calculated.

Paying attention to figure after adjusting and determining primary values algorithm started by measuring variations. In next step, the variations, compared to their old values and by considering this comparison results, one of the two rings or fine-tune of the operating point calculation selected. Figure 5, goes on to explain each block deal.

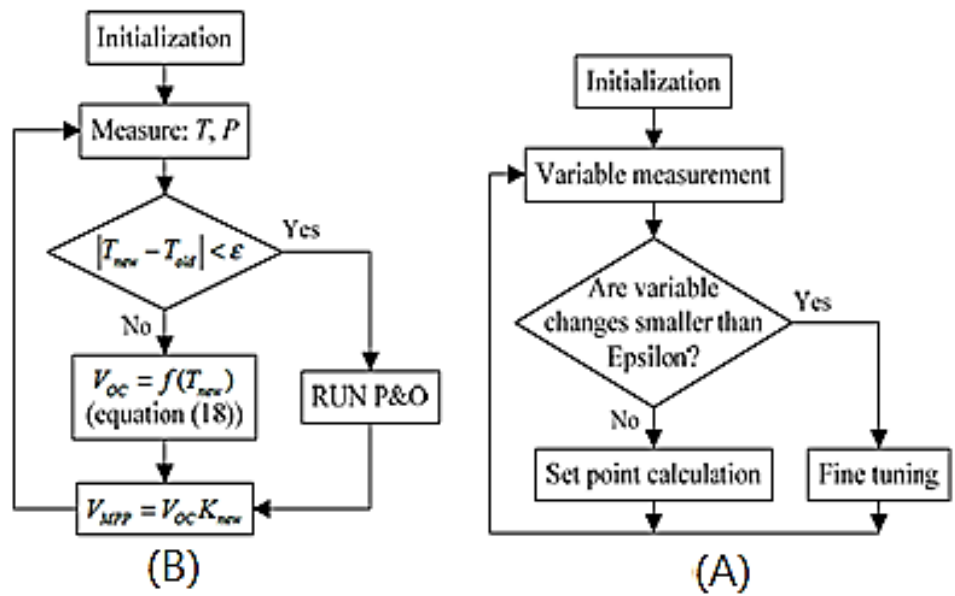

**Fig.3**. (a) The proposed method of general algorithm (b) the proposed method of algorithm

**The primary values:**

The primary values of photovoltaic system were classified in two groups.

**First group:**

Data provided by the manufacturer related to solar panel such as Voltage, current and power, peak power, open circuit voltage, short-circuit current in certain atmospheric conditions (temperature $25^{\circ C}$ and light intensity of 1,000 watts per square meter), temperature coefficient of short circuit current (KI) and Open voltage over temperature (KV).

**Second group:**

It consists a parameter that is used to trace power point, valued by operator and is the maximum voltage of the open circuit voltage (A), and the ratio of the maximum short circuit current. In proposed method, the values $V_{oc,n}$ and $K_v$ of first group and K value of second group has been determined as the primary values. The measurement variation: this variation consists of output variation (related to offline method) such as temperature, light intensity, voltage open and variables (related to system using online) such as current, voltage, power. In the proposed method, temperature and power are used.

Loop selecting: selecting the ring depends on the performance of each ring system conditions (steady state or transient). The algorithm calculates the operating point when the loop operated input changes are greater than a certain level and otherwise fine-tuning loop is executed. In proposed procedure if the temperature changes are greater than a calculation of loop certain point executed, the loop executed will be fine-tuned.

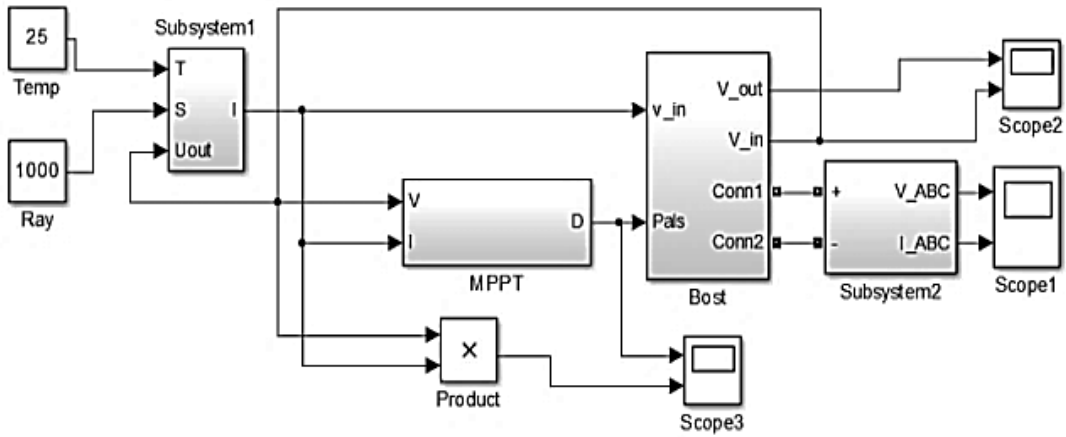

**Fig. 4.** Photovoltaic system Schematic with MPPT Grid mode

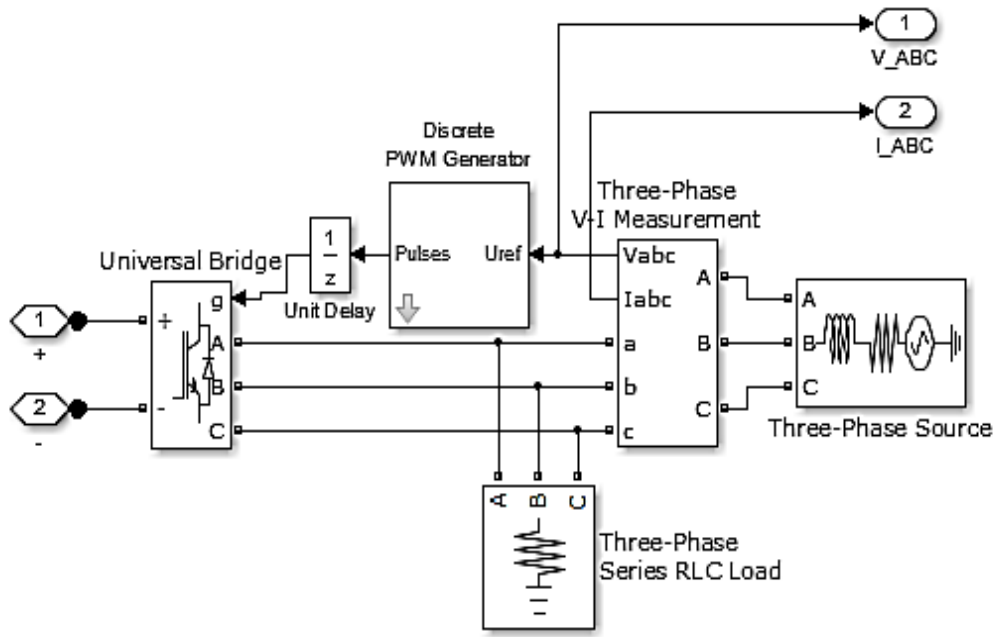

**Fig. 5.** Grid structure Schematic

49

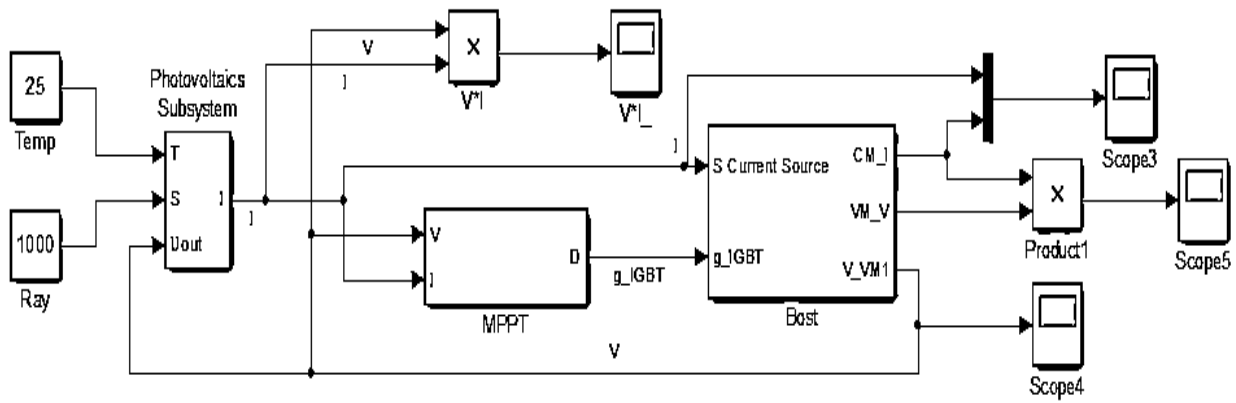

**Fig. 6.** Photovoltaic system Schematic with MPPT without Grid

Calculation of loop in certain point: in this loop by using primary datum and calculated values measured by the approximate amount of voltage, current or power of the maximum power point of the solar panel will be executed. This work implements based on offline methods and proposed algorithm work point calculation done based on the open circuit voltage. Then the following schematic diagram simulated, was presented in figures (4), (5) and (6).

## 4- Result and Disscution

The Simulink is a simulation tool with MATLAB software. The use of engineering simulation software Simulink is common, like many others, not confined to a particular application. By using Simulink the behavior of a system can be analyzed, without having to build it. As a result, engineers investigate the effect of disturbances or other input factors on performance by using Simulink in addition to the savings in cost and time to evaluate. Additionally, the systems Simulink gives this ability to appear response in the event of an input parameters change. Simulink is offered as a library in MATLAB provided by block simulation of this library as a block diagrams.

### 4-1- Non-Grid Mode

In order to investigate maximum power point, tracking ability presented current-voltage solar cell and then the specifications were used for maximum power point tracking state in which the techniques were provided.

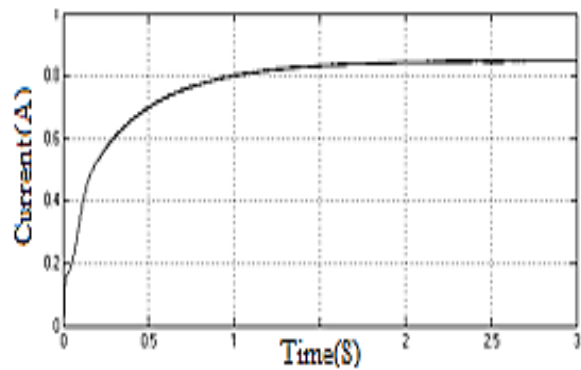

**Fig. 7.** The output current of photovoltaic cells with MPPT

Figure 8 shows the output current of photovoltaic by applying MPPT controller. Clearly, a significant reduction of the current turbulence was provided previously. At the beginning of the simulation, an output current of 25 mA was considered when the MPPT technique that entered the stream to any device connected to the associated photovoltaic cell destruction, MPPT technique triggered the output current process had gone smoothly, finally, it reaches to 0.83 mA, after 2 seconds, the current is fully proven with deleted disturbances.

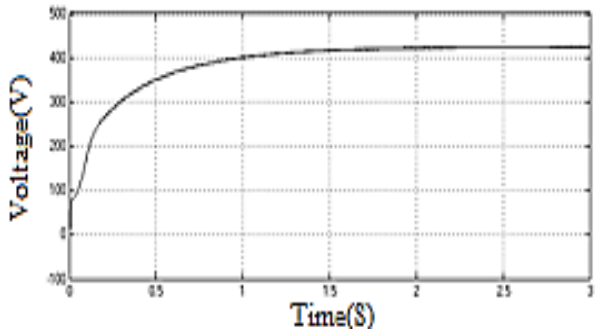

**Fig. 8.** The output voltage of photovoltaic cells with MPPT

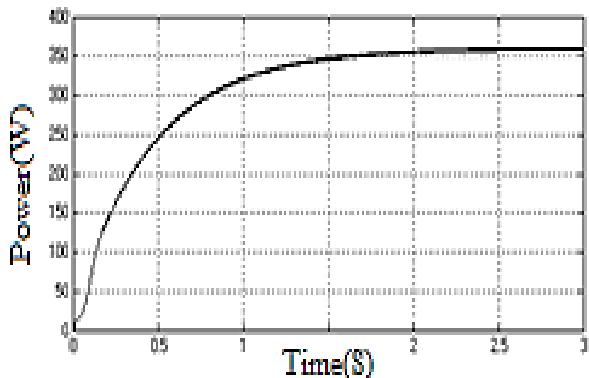

**Fig. 9.** Output power of photovoltaic cells with MPPT

The output voltage of photovoltaic cells with MPPT showed in fig 13 and adapting to presented figure clearly, a significant reduction of the voltage turbulence in proportional to voltage current without MPPT technique presented in figure 9. In figure 9, the photovoltaic cell is presented with MPPT technique. Amount of disturbances were reduced highly in power and 360 watts of maximum power MPPT technique is obtained. It must be noted that while working photovoltaic cell without exerting MPPT techniques equal to output power is in the range of 390 to 590 watts, 200-watt difference resulting disturbances can bring many of the techniques used in this study control output power in photovoltaic cell and omit obtained turbulences.

### 4-2- In Grid Mode

The second stage of simulation has been done to use photovoltaic cell with MPPT technique in grid mode. In this section, we present the simulation of MPPT technique in grid mode. In figure 19, output current MPPT techniques are presented in Grid Mode. Given that in the grid mode, the current has three phases. Therefore, we can observe the changes of flow in three-phase. Considering the figures presented in this field, it can be seen that at the beginning of the simulation the flow has set up turbulence and by passing time (0.7 s) this turbulence has been gone out and output current has not had turbulences in these three phases.

Fig. 11 shows the output voltage of the photovoltaic cell with MPPT techniques in Grid Mode. As it can be seen in the beginning of the simulation and in the range of 0.05 seconds, there is turbulence in the output voltage and by passing 0.05 second the turbulence will be reduced and this procedure will be constant in the end of simulation. In other words, MPPT techniques presented in this study are able to reduce three-phase flow turbulence effectively and total output voltage is equal to 30 volt.

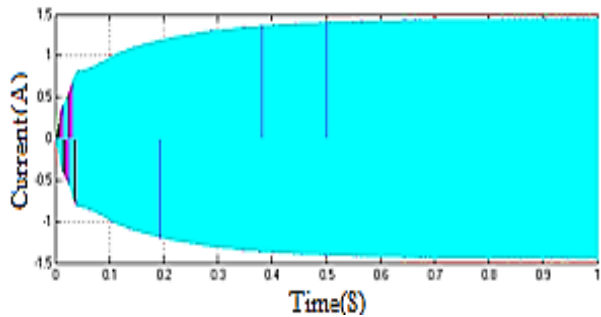

**Fig. 10.** The output current of the photovoltaic cell with MPPT techniques in Grid Mode

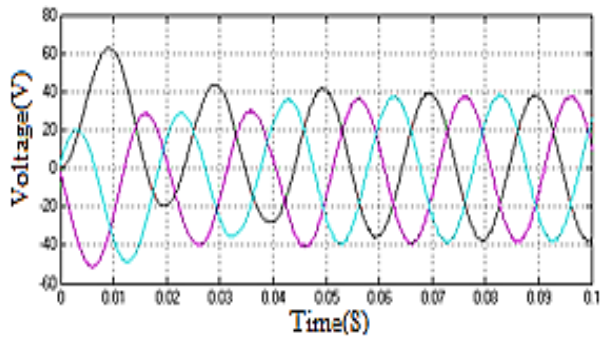

**Fig. 11.** The output voltage of the photovoltaic cell with MPPT techniques in Grid Mode

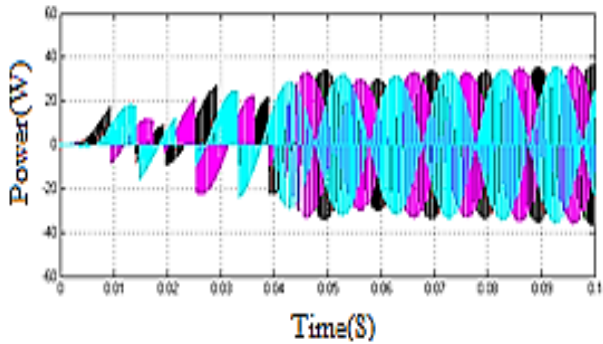

**Fig. 12.** The output power of the photovoltaic cell with MPPT techniques in Grid Mode

The fig. 12 shows output power of the photovoltaic cell with MPPT techniques in Grid Mode. Adapting to presented figure based on providing the output current and voltage of photovoltaic cell with MPPT techniques in Grid Mode, we can evaluate output power turbulence, because the resulting current of voltage turbulence has a more long range.

### 4-3- Total Harmonic Distortion

Total harmonic distortion (THD) is qualitative parameter and it indicates that a wave shape or signal to an extent close to a sine wave. THD value expressed as a percentage and the amount of THD is less than the sine wave with a better quality. Then in order to study further the output voltage of the photovoltaic cells with MPPT technology in Grid Mode, the THD reviews have been conducted presented in figure 12. Adapting to present diagram, THD value is equal to 9.56 percent that shows the best-proposed MPPT performance techniques.

### 5- Conclution

Nowadays the solar cell is being used in many issues. In present study, a new technique provided in order to track the maximum power point of photovoltaic cells and evaluated in both non-Grid connection and Grid connection. The obtained results of this study showed that the provided technique is able to control the output voltage and power of photovoltaic cell. The output voltage of photovoltaic cell THD value is equal to 9.56 percent and it is acceptable value by considering the existence standard.

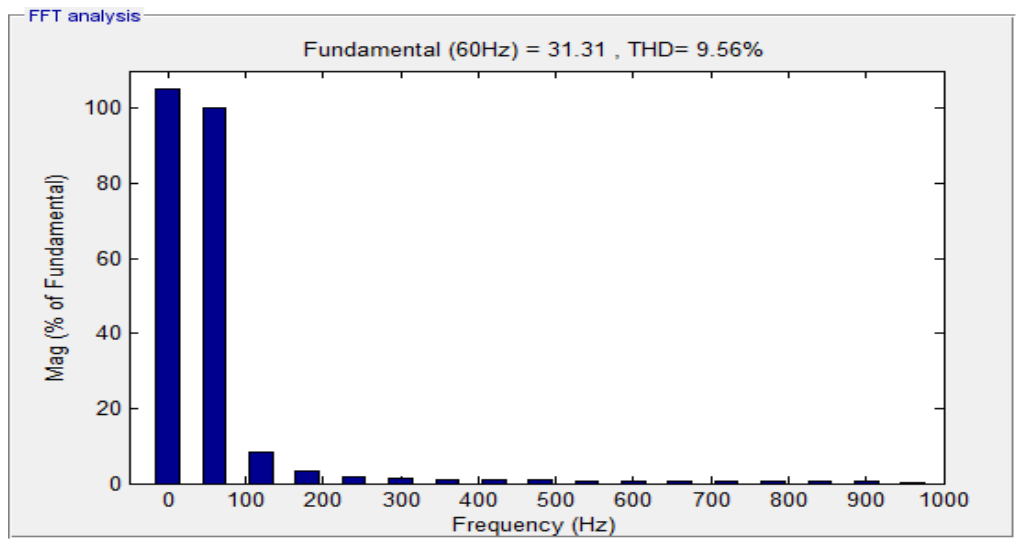

**Fig. 13.** The output voltage THD of the photovoltaic cell with MPPT techniques in Grid Mode (Hz)